\documentclass[11pt,twoside]{article}

\usepackage{asp2006}
\usepackage{epsf}
\usepackage{psfig}
\usepackage{lscape}
\usepackage[dvips]{graphicx}

\begin{document}
\title{Searching For Helical Magnetic Fields in Active Galactic Nuclei}   
\author{Mehreen Mahmud, Denise Gabuzda}   
\affil{Department of Physics, University College Cork, Cork, Ireland}    

\begin{abstract} 

Previously, a multi-wavelength (2cm, 4cm and 6cm) polarization study by Gabuzda, Murray and Cronin (2004) showed systematic Faraday Rotation gradients across the parsec-scale jets of several BL~Lac objects, interpreted as evidence for helical magnetic (B) fields~--- the gradients were taken to be due to the systematic variation of the line-of-sight B field across the jet. We present here new results for the parsec-scale Faraday Rotation distributions for eight additional BL~Lac objects, based on polarization data obtained with the Very Long Baseline Array (VLBA) at two wavelengths near each of the 2cm, 4cm and 6cm bands. The Rotation Measure (RM) maps for all these sources indicate gradients across their jets, as expected if these jets have helical B fields. Such gradients are also detected in the cores of several sources. This provides evidence that these gradients are present in appreciable regions of the jets and are not isolated phenomena. We also observe reversals in the RM gradient in the core region as compared to the gradient in the jet in at least three sources.

\end{abstract}


\section{Introduction}   

BL~Lac objects are  Active Galactic Nuclei (AGNs) characterized by strong and varying polarization at ultraviolet through radio wavelengths, which are observationally similar to radio-loud quasars in many respects, but display systematically weaker optical line emission. Very Long Baseline Interferometry (VLBI) polarization observations of BL Lac objects have shown a tendency for the magnetic (B) fields in their jets to be perpendicular to the local jet direction (Gabuzda et al 2000). It seems likely that many of these transverse B fields represent the ordered toroidal component of the intrinsic B fields of the jets, as discussed by Gabuzda (2007), see also references therein.

\section{Faraday Rotation}    
Faraday Rotation of the plane of linear  polarization occurs during the passage of an electromagnetic wave through a region with free electrons and a magnetic field with a non-zero component along the line-of-sight, and is due to the difference in the propagation velocities of the right and left-circularly polarized components of the wave. The amount of rotation is proportional to the density of free electrons $n_{e}$ multiplied by the line-of-sight magnetic field $B \cdot dl$, the square of the observing wavelength $\lambda^{2}$, and various physical constants; the coefficient of $\lambda^{2}$  is called the Rotation Measure (RM):

\begin{eqnarray}
           \Delta\chi\propto\lambda^{2}\int n_{e} B\cdot dl\equiv RM\lambda^{2}
\end{eqnarray}

Systematic gradients in the Faraday Rotation have been observed across the parsec-scale jets of several AGNs, interpreted as reflecting the systematic change in the line-of-sight component of a toroidal or helical jet B field across the jet  (Blandford 1993; Asada et al. 2002; Gabuzda, Murray \& Cronin 2004). Such fields would come about in a natural way as a result of the `winding up' of an initial `seed' field by the rotation of the central accreting objects (e.g. Nakamura, Uchida \& Hirose 2001: Lovelace et al. 2002). Determining how widespread this phenomenon is will be crucial for our understanding of AGN jets.

\section{Observations and Results}
Polarization observations of 37 BL Lac objects were obtained using the 10 25-m radio telescopes of the VLBA, at two wavelengths near the 2cm, 3.6cm and 6cm bands (a total of 6 wavelengths).  The data were obtained over 5  epochs (22 August 2003, 22 March 2004, 12 April 2004, 11 August 2004 and 2 September 2004), with 8-10 sources observed per epoch. The data were calibrated and imaged in the NRAO AIPS package using standard techniques. 
The uncertainties in the polarization angles shown take into account the rms noise in the Stokes Q and U maps and the uncertainty in the angle calibration. Results for nine of the BL Lac objects in which we have detected transverse RM gradients are presented here (Figs. 1-5). After matching the imaging parameters and beam sizes of the final images at  all the wavelengths, we constructed maps of the RM,  after first subtracting the effect of the integrated RM (presumed to arise in our Galaxy) from the observed polarization angles (Rusk 1988: Pushkarev 2001). 
Total-intensity contours with superimposed RM distributions and plots of polarization angle ($\chi$) versus wavelength ($\lambda$) squared are shown (Figs. 1-3). We observe an extended RM gradient in the jet of 0820+225 (Fig. 4); Gabuzda, Murray and Cronin (2004) also observed a gradient in the jet of this source, but somewhat closer to the core, possibly suggesting that the electron-density distribution has changed between our observations. The transverse RM gradients in the jets of 0003-066, 0256+075, 0716+714, 0735+178, 0954+658, 1156+295 and 2155-152 were detected by us for the first time. The gradient in 1749+096 (Fig. 3) is visible in the RM map of Zavala and Taylor (2004); the gradient in 0954+658 (Fig. 4) is confirmed by O' Sullivan \& Gabuzda, these proceedings. We also observe a reversal of the gradient in the core regions as compared to the jets in 0716+714 and 2155-152 (Fig. 5). See Mahmud et al, these proceedings, for an example of a source showing reversal of the gradient in the jet over time.

\begin{figure}
\centering
\includegraphics[height= 2.5 in]{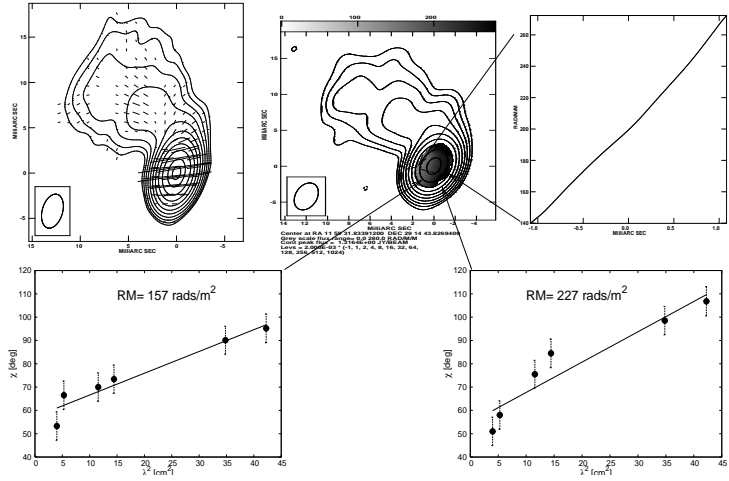}
\caption[Short caption for figure 1]{\label{Fig1.1}{RM map of 1156+295 (center) for epoch 22 August 2003,  with the overlaid Intensity (I) map at 5GHz. Also shown is an RM slice across the jet marked by the dotted line (top right). The I map of the source with the Electric Vectors overlaid shows a `spine-sheath' polarization structure (top left), indicative of a helical B field. The errors shown are 2 $\sigma$.}}
\label{Figure 1.1}
\end{figure}

\begin{figure}
\centering
\includegraphics[height= 2.0 in]{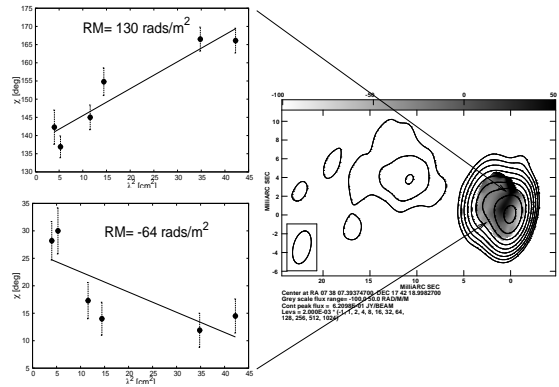}
\caption[Short caption for figure 1]{\label{Fig1.2}{RM map of 0735+178, epoch 11 August 2004, with the overlaid Intensity map. The errors shown are 1 $\sigma$.}}
\label{Figure 1.2}
\end{figure}

\begin{figure}
\centering
\includegraphics[height= 1.5 in]{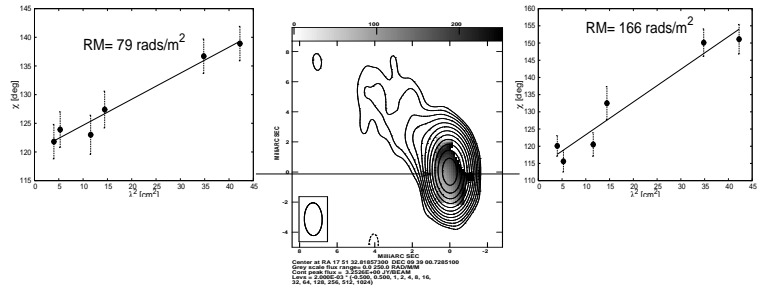}
\caption[Short caption for figure 1]{\label{Fig1.3}{RM map of 1749+096, epoch 22 March 2004, with the overlaid I map. The errors shown are 1 $\sigma$.}}
\label{Figure 1.3}
\end{figure}

\begin{figure}
 \begin{minipage}[t]{7.0cm}
 \begin{center}
\includegraphics[width=4.4cm,clip]{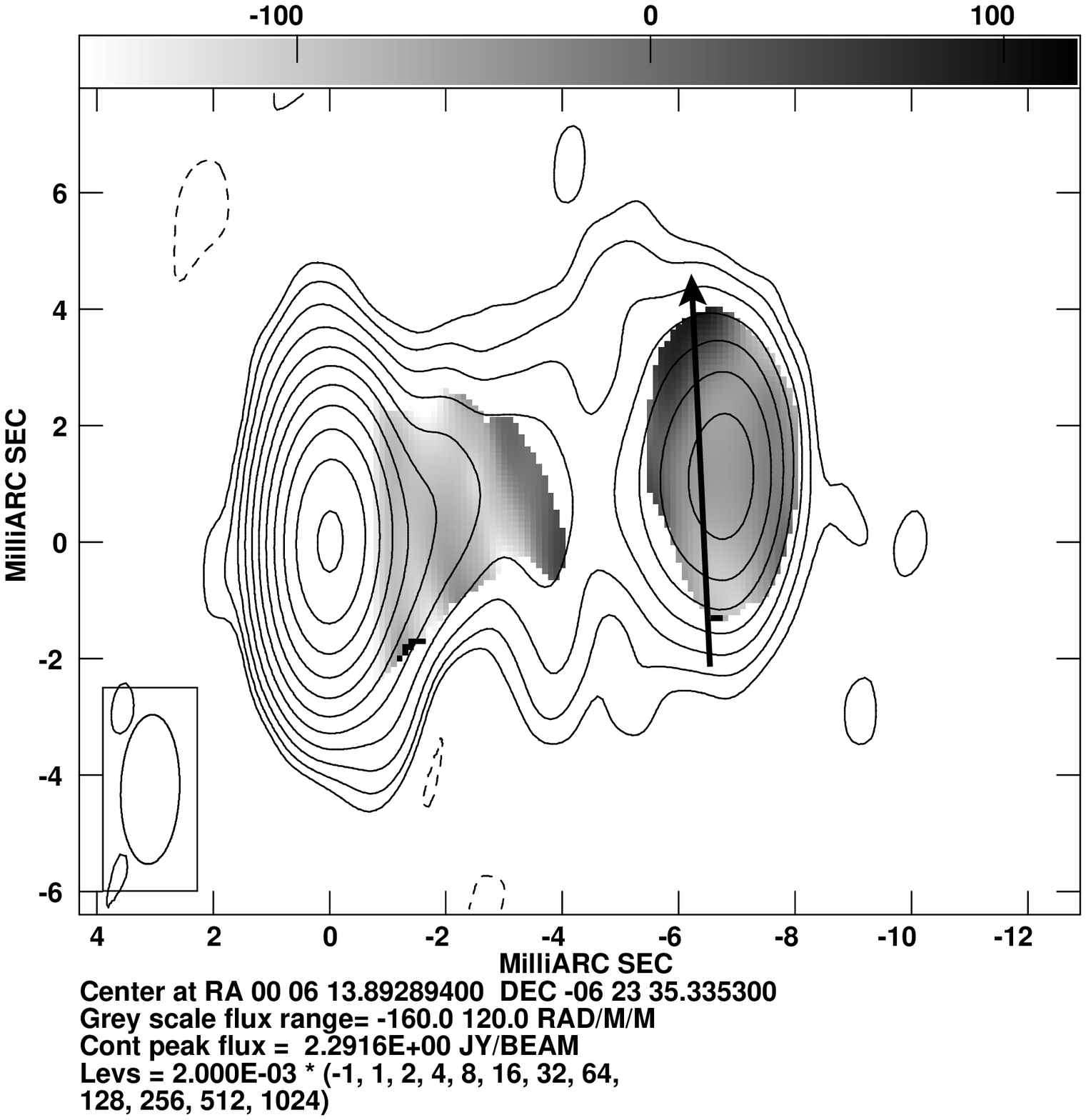}
 \end{center}
 \end{minipage}
 \hfill
 \begin{minipage}[t]{7.0cm}
 \begin{center}
 \includegraphics[width=3.4cm,clip]{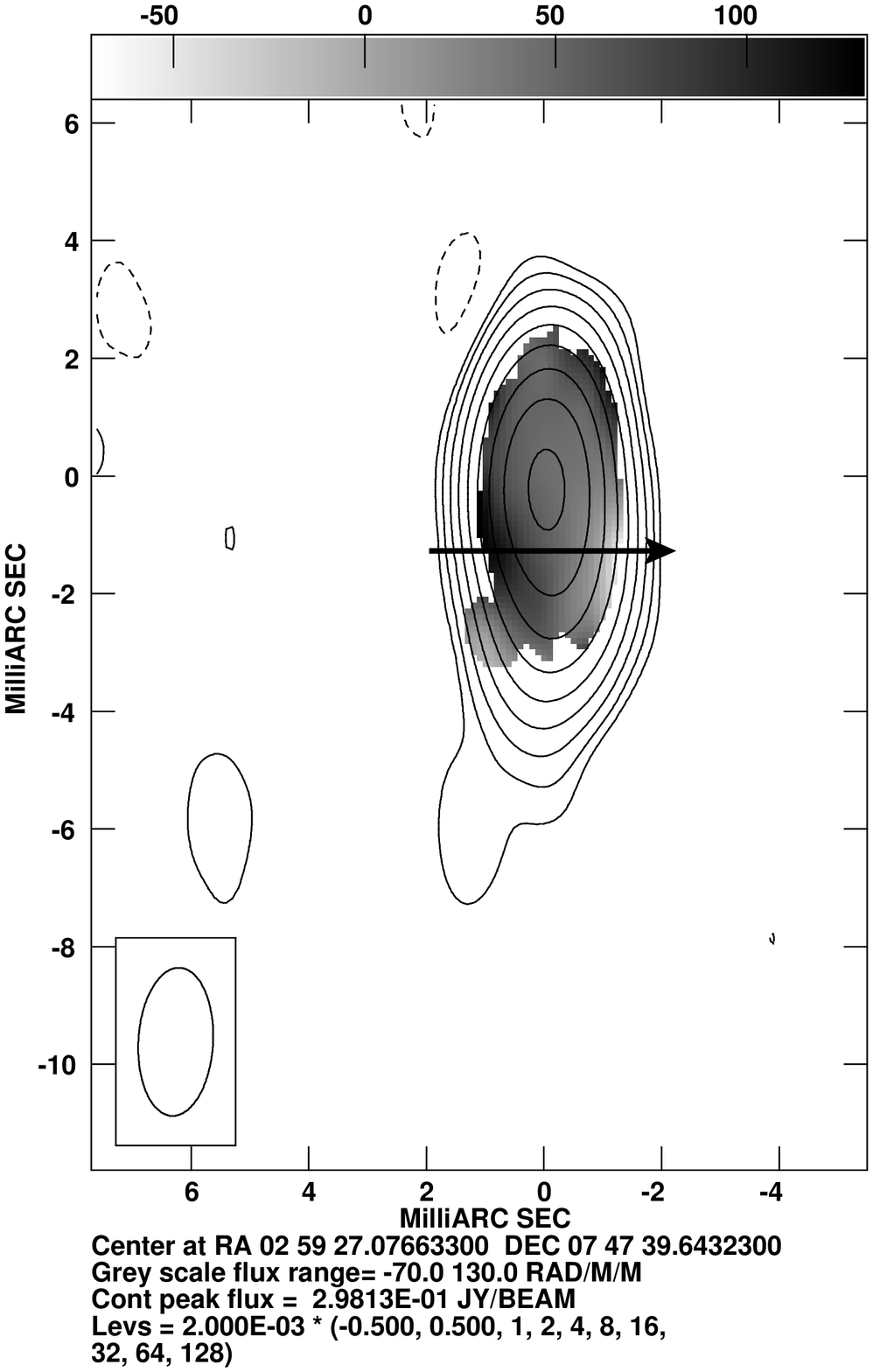}
 \end{center}
 \end{minipage}

 \begin{minipage}[t]{7.0cm}
 \begin{center}
 \includegraphics[width=4.9cm,clip]{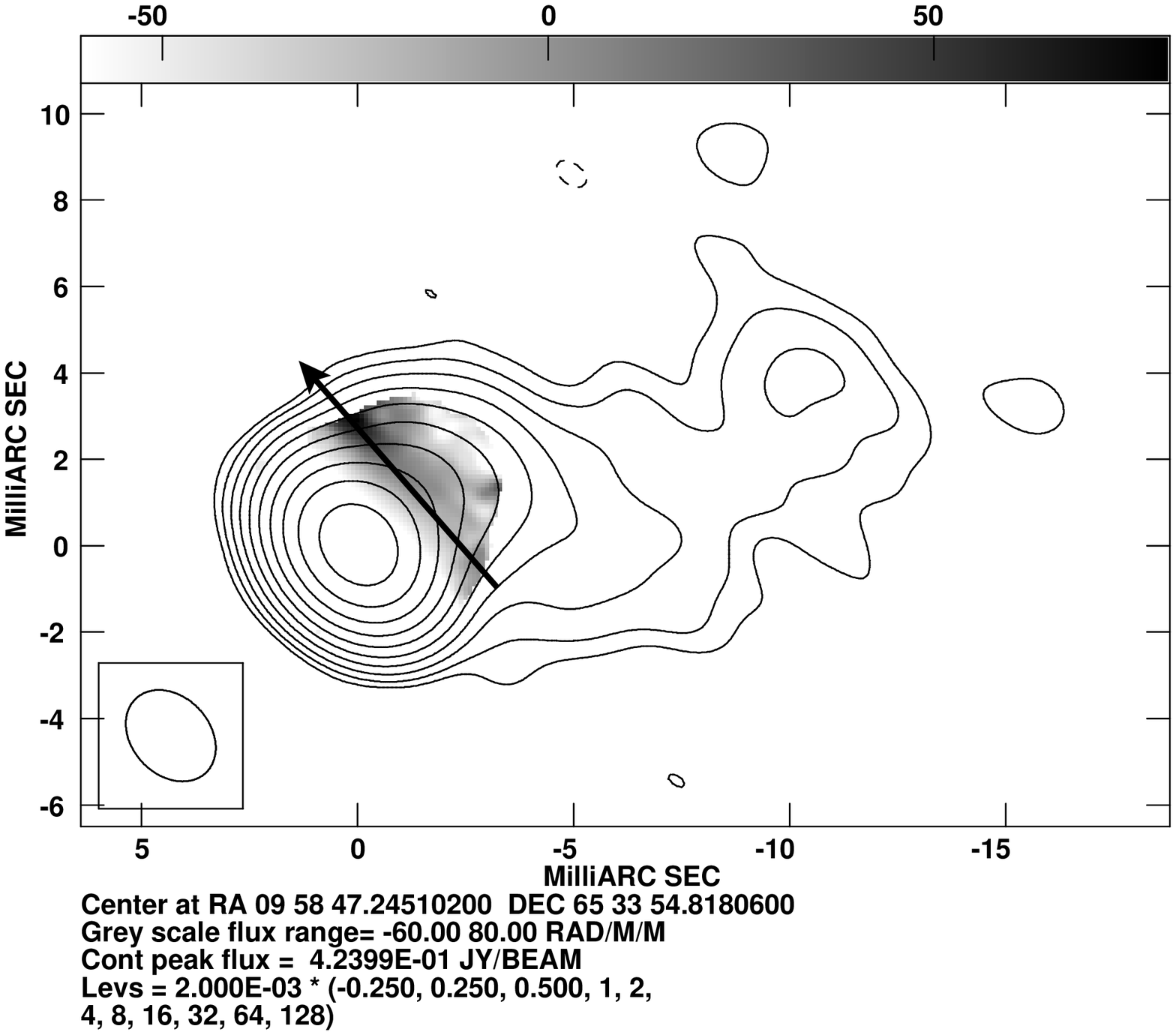}
 \end{center}
 \end{minipage}
 \hfill
 \begin{minipage}[t]{7.0cm}
 \begin{center}
 \includegraphics[width=3.9cm,clip]{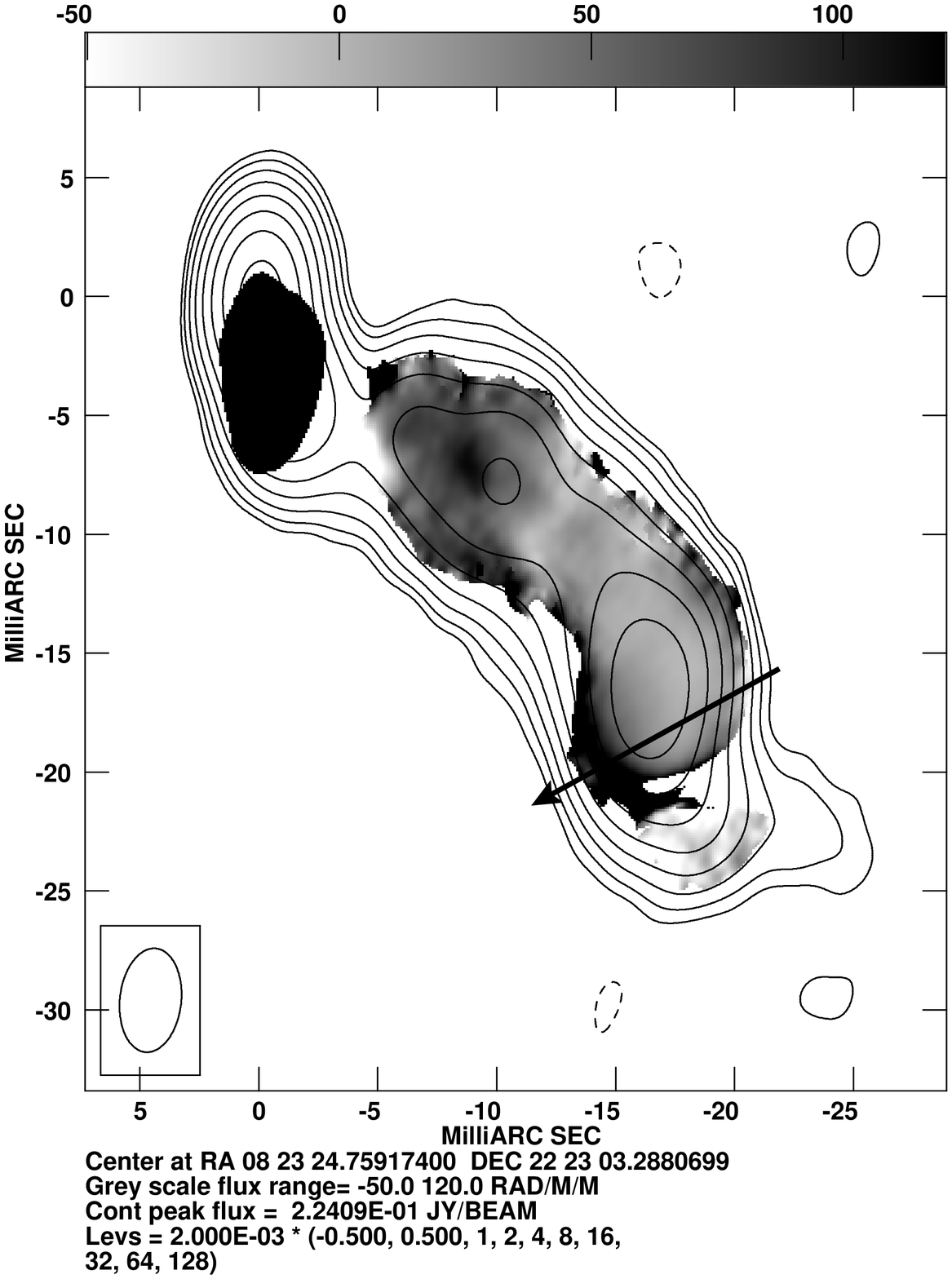}
 \end{center}
 \end{minipage}
\caption[Short caption for figure 1.4]{\label{labelFig1.4}{Shown above, clockwise, starting from top left, are RM maps with overlaid Intensity maps of 0003-066 (epoch 22 March 2004), 0256+075 (epoch 11 August 2004), 0820+225 (epoch 22 August 2003) and 0954+658 (epoch 12 April 2004), each showing transverse RM gradients, shown by the arrows.}}
\end{figure}

\begin{figure}
 \begin{minipage}[t]{4.3cm}
 \begin{center}
 \includegraphics[width=3.1cm,clip]{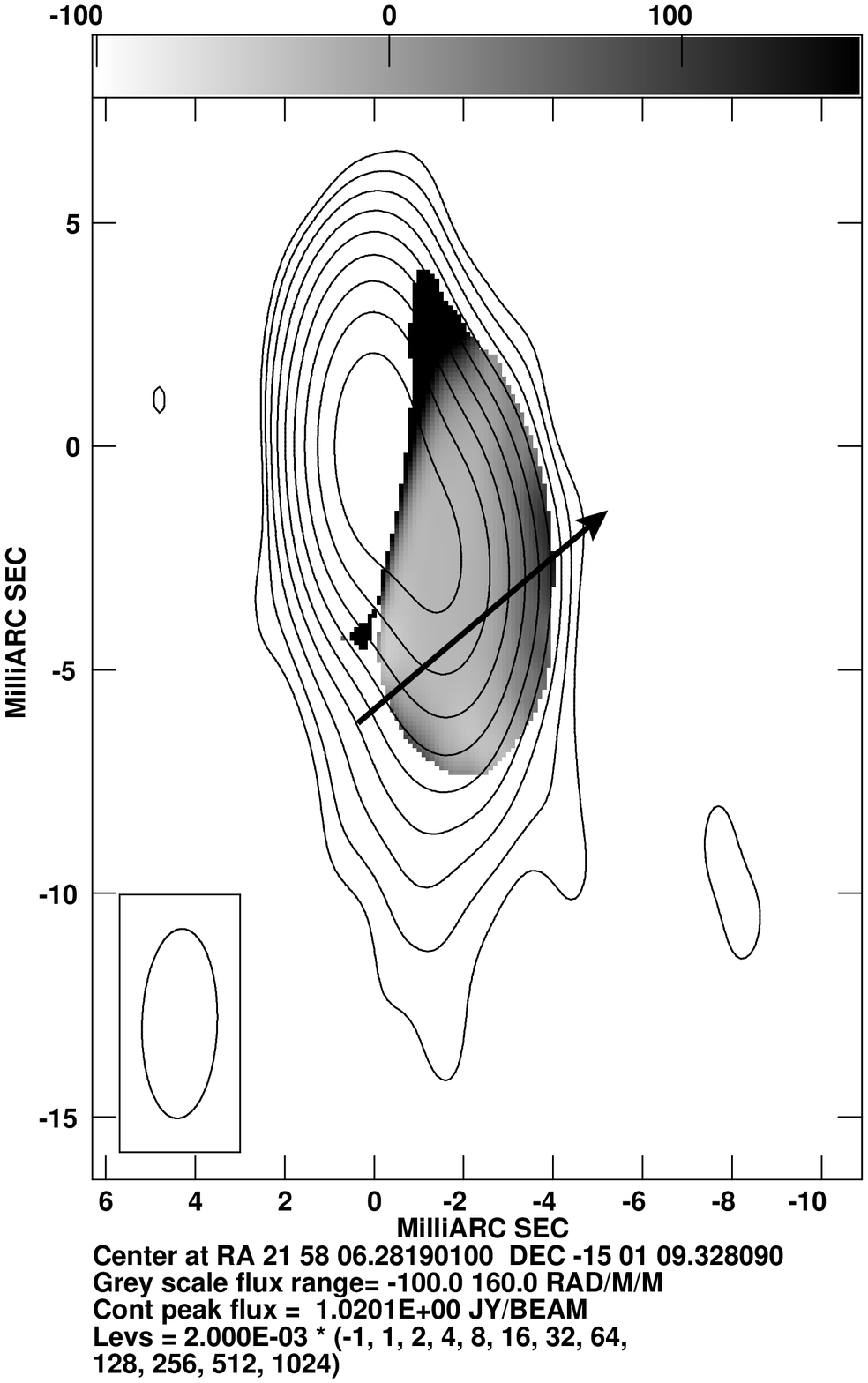}
 \end{center}
 \end{minipage}
 \hfill
 \begin{minipage}[t]{4.3cm}
 \begin{center}
 \includegraphics[width=3.1cm,clip]{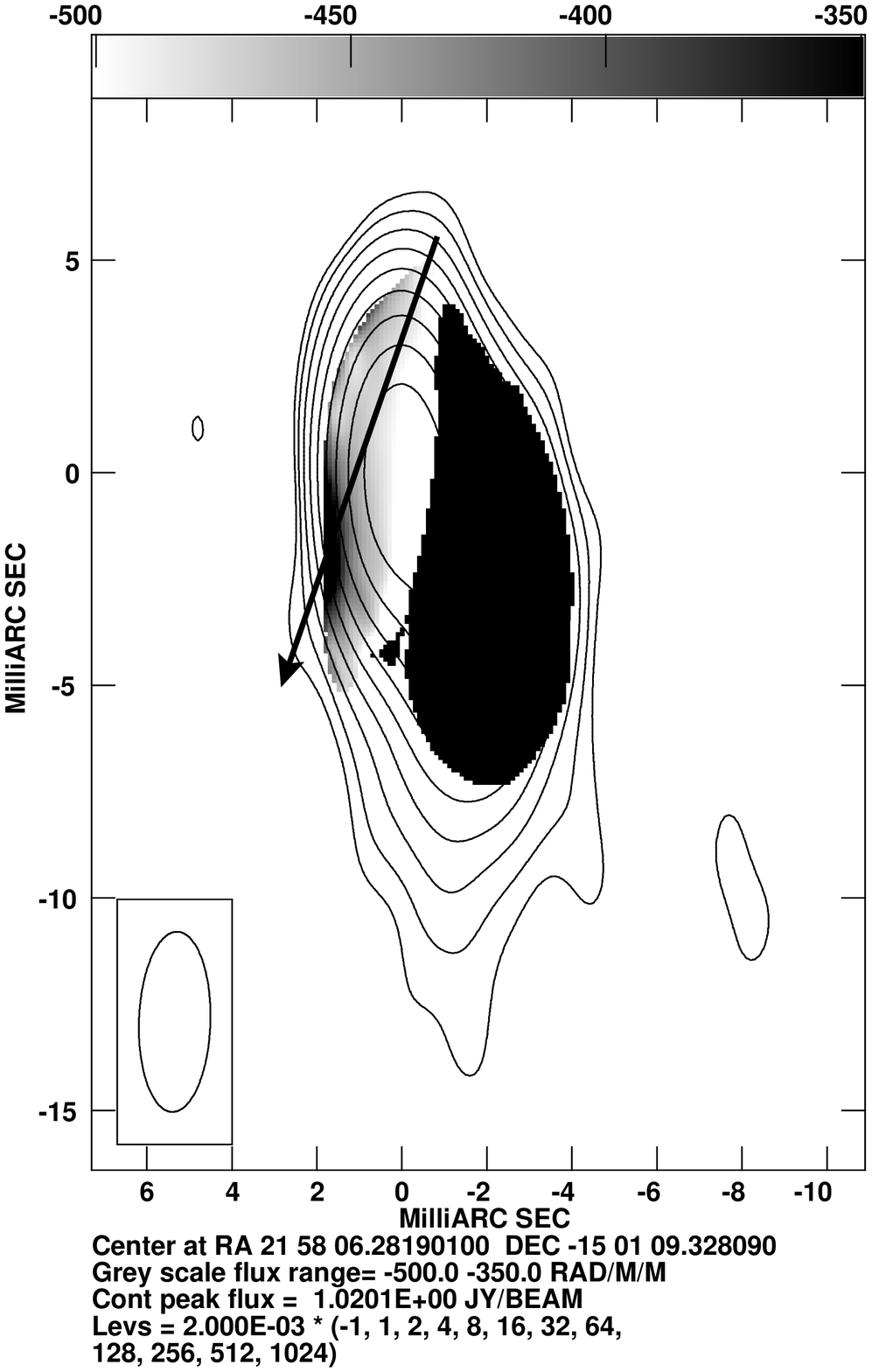}
 \end{center}
 \end{minipage}
\begin{minipage}[t]{4.3cm}
 \begin{center}
 \includegraphics[width=3.1cm,clip]{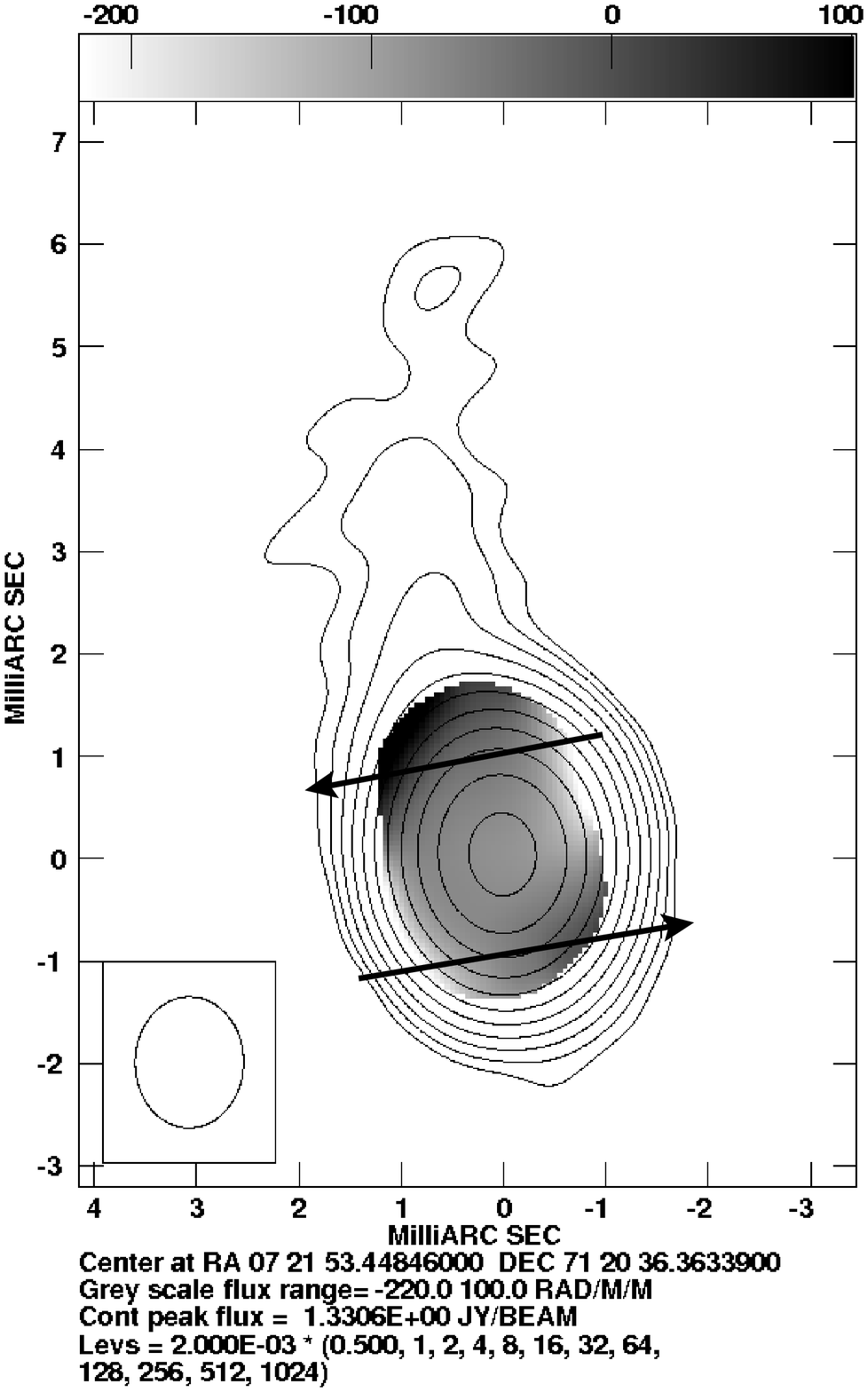}
 \end{center}
 \end{minipage}
\caption[Short caption for figure 2]{\label{labelFig2} {RM maps for 2155-152 and 0716+714 (epoch 22 March 2004); the first map detailing the gradient in the core region of 2155-152, the second map showing the gradient in the jet of 2155-152, opposite in direction to the core (the arrows show the RM gradient in each region), and the third map showing the gradient in the core and jet of 0716+714.}}
\end{figure}

\section{Discussion}
The presence of transverse RM gradients has been confirmed for the jets of 0820+225 (Gabuzda, Murray \& Cronin 2004) and 1749+096 (Zavala \& Taylor 2004), and new transverse RM gradients have been detected in the jets of seven other objects. This clearly demonstrates that this is a fairly widespread phenomenon which, in turn, suggests that helical B fields are common in AGN jets. Furthermore, we find evidence for evolution in the RM distribution in 0820+225. The `spine-sheath' polarization structure (transverse magnetic field vectors in the central jet region, with longitudinal magnetic field vectors hugging the edges of the jet) in 1156+295 is consistent with this jet having a helical magnetic field (Lyutikov et al 2005). Our observations have also revealed new features not previously observed, such as reversals in these gradients with distance from the core (Fig. 5) and with time (Mahmud et al, these proceedings). The mechanisms behind these phenomena are not yet fully understood but may provide information about the intrinsic magnetic field configuration that is `wound up' by the rotation of the central black hole. 
Some of the results presented here are analyzed in more detail by Gabuzda et al (2007).

\acknowledgements 
This publication has emanated from research conducted with the financial support of Science Foundation Ireland. The National Radio Astronomy Observatory is operated by Associated Universities Inc.


\end{document}